# Model Contribution Rate Theory: An Empirical Examination

**Vincil Bishop III**[†]
Department of Systems Engineering
Colorado State University Alumnus
Vincil.Bishop24@alumni.colostate.edu

**Steven Simske**
Department of Systems Engineering
Colorado State University
Steve.Simske@colostate.edu

## Abstract

The paper presents a systematic methodology for analyzing software developer productivity by refining contribution rate metrics to distinguish meaningful development efforts from anomalies. Using the Mean-High Model Contribution Rate (mhMCR) method, the research introduces a statistical framework that focuses on continuous contributions, mitigating distortions caused by tool-assisted refactoring, delayed commits, or automated changes. The methodology integrates clustering techniques, commit time deltas, and contribution sizes to isolate natural, logical work patterns and supports the accurate imputation of effort for contributions outside these patterns. Through empirical validation across multiple commercial repositories, the mhMCR method demonstrates enhanced precision in productivity measurement in identifying sustained developer activity. The findings provide actionable insights for optimizing team performance and workflow management in modern software engineering practices.

## 1 Introduction

In the increasingly complex landscape of collaborative software development, accurately quantifying developer contributions has become a significant challenge. Modern workflows often integrate AI-assisted tools, systematic refactoring, and other automated processes that obscure the distinction between genuine development efforts and machine-driven or extraneous changes. The goal of this research is to refine the accuracy of model contribution rates—key metrics that represent sustained, meaningful developer activities—and enhance their use in the Contribution Rate Imputation Method (CRIM) [1].

This paper focuses on defining and empirically validating a methodology for identifying "model contributions," which serve as the foundation for reliable productivity metrics. By differentiating model contributions from non-model or unnatural contributions, such as mass refactoring or wholesale code duplication, the proposed approach aims to improve imputation methods and provide a precise, scalable framework for analyzing software team performance. The result is a refined model that isolates true developer efforts, ensuring that productivity metrics align with real-world contributions.

---

[†] Not related to work at Amazon, Inc.

This work builds on previous research, including the "Time-Delta Method for Measuring Software Development Contribution Rates" dissertation, from which it heavily borrows concepts and methodologies [2].

## 2 Background and Related Work

Various methodologies have been employed to explore the analysis of developer productivity through commit time deltas and contribution sizes, leveraging clustering techniques to derive insights from commit histories and contribution metrics. This synthesis of background and related works will delve into the relevant studies and methodologies that contribute to understanding developer productivity through these analytical lenses.

One prominent approach involves the utilization of clustering techniques to analyze commit messages and their associated metadata. Yang et al. [3] conducted a study that employed natural language processing (NLP) to extract key information from commit messages, which was subsequently used for clustering and classifying commits. Their findings indicated a minimal feature reduction percentage of 0.83%, yet they achieved a notable F1-score, suggesting that even small improvements in feature extraction can enhance the clustering process. This study underscores the importance of commit messages as a rich source of information that can be leveraged to analyze developer productivity.

In a broader context, Liu et al. [4] highlighted the critical role of clustering techniques in software component analysis. They pointed out that traditional clustering methods often rely on a single metric to measure similarity, which may not capture the complexity inherent in software systems. This limitation suggests that a multifaceted approach, incorporating multiple metrics such as commit time deltas and contribution sizes, could yield more accurate clustering results. By integrating various metrics, researchers can better understand the relationships between different components and their contributions to overall developer productivity.

Barman et al. [5] further explored clustering techniques specifically tailored for software engineering. Their research focused on the effectiveness of a Genetic Algorithm-based Software Modularization Clustering (GASMC) method applied to real-world software module clustering problems. The results demonstrated that their proposed method significantly improved clustering outcomes, thereby facilitating a better understanding of how different software modules contribute to overall productivity. This study emphasizes the potential of advanced clustering techniques to enhance the analysis of developer contributions.

Jeon et al. [6] introduced Githru, a visual analytics tool designed to analyze software development history through Git metadata. Their approach included a clustering step that allowed for the examination of commit trends over time, providing insights into developer activity patterns. By visualizing clusters of commits, developers and project managers can identify periods of high productivity or potential bottlenecks, thereby informing resource allocation and project planning.

The concept of untangling commits to improve the clarity of commit histories has also been explored. Partachi et al. [7] presented Flexeme, a technique that utilizes a multi-version Program Dependency Graph to capture the relationships between different program versions. By annotating data flow edges with names and lexemes, their approach facilitates the identification of distinct contributions made by developers. This untangling process is crucial for accurately assessing developer productivity, as it ensures that each commit reflects a coherent set of changes related to a specific task.

In addition to these studies, methodologies that focus on fine-grained code changes have emerged as vital tools for analyzing developer productivity. Yi et al. [8] proposed a dynamic delta refinement approach to produce smaller semantic history slices, which can help in understanding the significance of changes in relation to target functionalities. By refining the analysis of commit deltas, developers can gain insights into the impact of their contributions on the overall software system, thereby enhancing productivity assessments.

The Delta Maintainability Model (DMM) introduced by Biase et al. [9] offers a framework for measuring the maintainability of fine-grained code changes. This model allows for the evaluation of individual commits, providing a detailed understanding of how specific changes affect the maintainability of the software. By quantifying the maintainability of commits, developers can prioritize their contributions based on their impact on the software's long-term viability.

The integration of clustering techniques with machine learning has also shown promise in predicting developer productivity. Gupta and Kang [10] explored fuzzy clustering approaches for predicting the severity of faults in software systems. Their findings suggest that clustering can be effectively utilized to categorize developer contributions based on their potential impact on software quality, thereby informing decision-making processes related to resource allocation and project management.

## 3 Theory

### 3.1 Model Contribution Rate Theory

If *model contributions* associated with continuous and typical work sessions can be identified and quantified, the resulting contribution rate may be considered the *model contribution rate*. This baseline rate can then serve as a reference point for comparison against other types of contributions, allowing for the analysis of deviations or patterns in contribution behavior.

### 3.2 Contribution Classes

This section describes broad classes observed in contributions that underpin model contribution rate theory.

#### 3.2.1 Natural Contributions

Natural Contributions refer to code changes that stem from logical, intentional efforts by developers, whether authored manually or with the assistance of AI tools. These contributions directly reflect the development process and are pivotal in determining productivity baselines for individuals and teams.

Key characteristics of natural contributions include their origin in continuous work sessions that are not interrupted by unrelated tasks or systematic, non-logic-related operations such as mass refactoring. Natural contributions are distinct from unnatural contributions, which may arise from automated processes or wholesale code duplication, neither of which reflects the effort or logical progression inherent to software development.

Identifying natural contributions involves clustering commit data using metrics such as commit time delta (CTD) and contribution size. Contributions that occur within a defined CTD range, indicative of continuous and logical work, are classified as model contributions. This process ensures a focus on developer-driven efforts, enabling more accurate measurement of productivity and better insights into team dynamics.

Such refined categorization aids in excluding anomalies and enhances the precision of contribution metrics, providing a robust framework for analyzing and benchmarking development performance.

*Machine Generated / AI Assisted Code*

In this context, AI-assisted contributions, when employed to produce complete, functional code, are considered natural. The reasoning is that such contributions enhance a developer's productivity by accelerating logical software development. For instance, if an AI tool generates new logic based on a developer's direction, the resulting contribution is treated as a valid and productive effort. This approach supports the notion that AI-driven enhancements are integral to modern development workflows and should be credited to developers as part of their natural productivity baseline.

This exploration takes the position that AI-assisted code changes should be credited to the developer as productive contributions, even though the changes are machine-generated. The rationale is that AI-generated code represents complete and functional logic, significantly boosting developer productivity. Consequently, these machine-assisted contributions reflect genuine productivity and can be integrated into the developer's—and their team's—productivity baseline.

The objective is to distinguish between productive AI-assisted contributions and anomalous machine-generated contributions. Anomalies occur when machine-generated changes do not involve the addition of meaningful logic but instead result from systematic operations, such as automated refactoring. These refactoring operations produce large textual changes that lack logical progression and are unrelated to actual software development efforts. By identifying such anomalies, it becomes possible to separate meaningful contributions—whether human-authored or AI-assisted—from irrelevant or inflated changes that distort productivity metrics.

By contrast, contributions resulting from automated refactoring tools or systematic changes that do not add new logic paths to the codebase are classified as unnatural contributions. These include operations like renaming variables en masse or reformatting code, which, despite generating textual changes, do not align with the logical progression of a software project.

### 3.2.2 Unnatural Contributions

Unnatural contributions in software development refer to code changes that deviate from the typical patterns of logical and intentional human programming effort. These contributions often emerge from automated processes, such as tool-assisted refactoring or wholesale copying of code, which, while technically modifying the codebase, do not reflect the logical progression or creative problem-solving inherent in traditional development workflows.

*Types of Unnatural Contributions*

1. **Mass Refactoring:** Involves tool-generated changes applied systematically across the codebase, such as renaming variables, reformatting, or restructuring code for consistency. While these operations improve code maintainability, they lack the logical development effort typically required to introduce new functionality or resolve complex problems.
2. **Copied Code:** Occurs when large portions of code are imported wholesale into a repository. These contributions are particularly challenging to detect, as they can appear as significant additions without providing insight into the actual effort or problem-solving process behind their creation.

*Identifying Unnatural Contributions*

Unnatural contributions can be detected through clustering techniques that analyze metrics such as commit time delta (CTD) and contribution size. For instance:

- **Low CTD, High Contribution:** Contributions with short intervals between commits but large changes often indicate automated processes like IDE-assisted refactoring.
- **High CTD, Low Contribution:** Delayed commits with minimal changes may reflect copied code or non-continuous work sessions.

*Implications of Unnatural Contributions*

Inclusion of unnatural contributions in productivity metrics can distort analyses, overestimating a developer's effort and contribution rate. For example:

- Mass refactoring commits may appear as high-output periods without reflecting genuine productivity.
- Copied contributions can inflate metrics without representing actual time or logical effort spent on development.

By identifying and excluding these contributions, productivity assessments can better align with actual developer effort, ensuring metrics are more representative of logical, creative, and sustained work sessions. This refined categorization supports more accurate performance evaluations and aids in understanding team dynamics and workflow patterns.

### 3.2.3 Quick Remedy Commits

"Quick remedy commits," as presented by Wen, et al., are defined as commits made by developers shortly after a prior commit, aimed at remedying issues introduced or changes omitted in the previous commit [11]. These commits serve as a follow-up to:

1. **Fix Errors or Omissions:** Address errors like broken references or incomplete logic due to an oversight in the earlier commit. For instance, developers may fail to propagate changes across all relevant code components, leading to quick remedy commits to fix those issues.
2. **Finalize Work:** Implement changes left unfinished in the prior commit, such as completing the addition of test cases or finalizing a refactoring effort.
3. **Improve Previous Commit:** Refactor or optimize code to improve clarity or performance after initially committing less optimal changes.

*Impact on Characterizing "Normal" Work Sessions*

Quick remedy commits may not align with typical patterns of cohesive or planned development. They often emerge from mistakes, omissions, or adjustments realized in hindsight. When characterizing "normal" work sessions, these commits can appear out of character because:

- They may introduce noise in the data used to analyze development practices, as their reactive nature deviates from the planned or strategic workflow.
- They can artificially inflate metrics like commit frequency or bug-fix rates if not properly identified and accounted for in the analysis.
- They highlight areas where tooling or processes might be insufficient, such as lack of automated checks that could prevent omission errors.

For accurate characterization of "normal" sessions, quick remedy commits should be treated distinctly. Evaluation process should account for their unique nature by identifying and possibly excluding them from datasets focusing on cohesive and intentional development behaviors.

### 3.2.4 Model Contributions

Model contributions represent the essence of typical, sustained, and logical development efforts in software engineering. They are a conceptual category of contributions that reflect the continuous, intentional work patterns of developers. These contributions align closely with natural development workflows, capturing the kind of changes that developers make when they are actively engaged in solving problems, implementing features, or refining software in meaningful ways.

The theoretical significance of model contributions lies in their role as a benchmark for productivity measurement within the Model Contribution Rate Theory. By focusing on these contributions, the theory establishes a baseline that isolates genuine development efforts from anomalies or extraneous operations, such as automated refactoring or large-scale, non-contextual changes. This baseline provides a reference point to understand and evaluate the quality and intensity of development activity across teams or individuals.

Model contributions are central to understanding the dynamics of Model Contribution Rate Theory. They offer a framework for differentiating authentic efforts from distortions caused by unnatural contributions. By identifying these contributions, the theory not only ensures a more accurate assessment of productivity but also supports deeper insights into team performance and workflow optimization, aligning metrics with real-world developer impact.

### 3.2.5 Anti-Model Contributions

Converse to model contributions are **Anti-Model Contributions (AMCs).** These contributions refer to commits that deviate from continuous, logical work patterns typically seen in productive coding sessions. These contributions often occur in scenarios where the **Commit Time Delta (CTD)**—the time elapsed between consecutive commits—does not correspond to uninterrupted or sustained development [2]. Such deviations can skew metrics, leading to an over- or underestimation of actual developer effort.

AMCs typically exhibit characteristics like:

1. **Low CTD but High Contribution**: These may indicate rapid, tool-assisted operations, such as mass refactoring, rather than deliberate and logical development.
2. **High CTD but Low Contribution**: Often suggestive of delayed commits or negligible updates, possibly reflecting minimal effort during the interval.

The presence of AMCs can distort **Contribution Rate (CR)** measurements, particularly when evaluating developer productivity. For example, mass refactoring might register as a large contribution but does not represent creative or logical software development. Similarly, delayed commits may exaggerate the perceived duration of development efforts.

### 3.3 Contribution Rate Imputation

The Contribution Rate Imputation Method (CRIM) focuses on imputing effort during unobserved work periods using historical commit data, combining metrics like contribution size (e.g., Levenshtein distance) with a model contribution rate derived from observed patterns. It enables estimation of time spent on contributions, particularly where direct tracking is unavailable, and accommodates variability in effort due to task complexity and context [1]. A primary motivation for refining model contribution rate methods is to improve the accuracy of contribution rate imputation.

#### 3.3.1 Unlikely Resolved Effort

The concept of **Unlikely Resolved Effort (URE)** provides a quality assurance mechanism for evaluating the plausibility of contribution rates derived from commit data. It operates on the assumption that a contributor is unlikely to sustain more than a practical threshold of resolved effort—such as working eight hours a day continuously on a single contribution. Contributions exceeding this threshold signal potential inaccuracies in the estimated contribution rate or inconsistencies in the imputation method.

In practice, the URE Theory evaluates the **Resolved Effort Hours (REH)**, derived from an imputed contribution rate and metrics such as commit size or time delta (CTD). If the REH exceeds plausible limits (e.g., eight hours per day), the contribution rate is flagged as overestimated, and the associated contribution is classified as exhibiting URE.

For practitioners observing the data, the rate of URE within a dataset serves as a proxy for assessing the reliability of contribution rate models. A higher incidence of URE suggests that the model's imputation or clustering process may lack precision, potentially distorting the productivity metrics.

This theory aids in model validation by introducing a sanity check, allowing practitioners to compare competing contribution rate models (e.g., **Mean Contribution Rate (MCR)** vs. **High Mean Contribution Rate (HMCR)**). Models yielding fewer URE instances are interpreted as offering greater accuracy in aligning imputed metrics with realistic developer effort.

URE also facilitates iterative refinement of imputation techniques by providing feedback on whether adjustments to CTD thresholds, clustering methods, or statistical measures improve the quality of contribution estimates. This ensures productivity assessments are both credible and actionable in real-world development contexts.

## 4 Mean-High Model Contribution Rate Method

### 4.1 Procedural Framework

The procedural framework of the Mean-High Model Contribution Rate (mhMCR) method is designed to systematically identify and quantify meaningful contributions in software development. This framework combines statistical techniques with logical inference to ensure the separation of natural and productive contributions from anomalous or irrelevant ones.

**Step 1: Initial Filtering of Commits**

The process begins with isolating commits that fall within the **Model Commit Time Delta Range (MCTDR)**. This range is predefined based on empirical data that identifies patterns associated with natural and continuous work sessions. The commits in this range, known as Model Contribution Candidates (MCC), serve as the initial dataset for further analysis. The MCTDR acts as a sieve, filtering out contributions that are either too rapid (indicative of automated tools) or too delayed (suggesting interruptions or irregular workflows).

**Step 2: Statistical Outlier Removal**

Next, the Interquartile Range (IQR) method is applied to the MCC dataset. By calculating the first (Q1) and third quartiles (Q3) and the interquartile range (IQR = Q3 - Q1), contributions that fall outside the bounds of $Q1 - 1.5 \times IQR$ and $Q3 + 1.5 \times IQR$ are identified as outliers. These outliers, which may represent either exceptionally low or high contributions, are excluded from further analysis to maintain the integrity of the dataset. The recalculated IQR ensures a refined focus on typical development patterns.

**Step 3: Classification of Contributions**

The remaining commits are classified based on their placement within the revised quartile ranges:

- **Q4 (High Contribution Range):** Contributions in this quartile are labeled as **Model Contributions (MC)**, representing normal productivity and sustained, logical development efforts.
- **Q1, Q2, and Q3:** These contributions, termed **Disqualified Model Contribution Candidates (dMCCs)**, do not meet the criteria for MCs and are excluded from the calculation of the model contribution rate. These contributions are also labeled as anti-model contributions for the purpose of contribution rate imputation.

**Step 4: Calculation of Model Contribution Rate (MCR)**

The model contribution rate is calculated by taking the mean of the contribution rates of the MCs identified in Q4. This focus on the highest quartile emphasizes peak productivity while excluding contributions that could distort the measurement due to their anomalous nature. The resulting MCR provides a robust benchmark for evaluating individual or team productivity.

**Conceptual Rationale**

The use of the highest quartile (Q4) aligns with the hypothesis that sustained, normal work sessions are indicative of meaningful developer contributions. By excluding dMCCs and outliers, the framework ensures that the calculated MCR reflects only the most representative efforts. This approach mitigates the impact of noise introduced by unnatural contributions, such as mass refactoring or delayed commits, enhancing the reliability of productivity metrics.

**Practical Implications**

This procedural framework is designed for scalability and adaptability in diverse software development environments. By leveraging commit time deltas and contribution sizes, it provides a data-driven methodology for organizations seeking to measure and improve developer performance. The mhMCR method can also serve as a foundation for advanced analytics, enabling deeper insights into team dynamics, workflow optimization, and the effects of AI-assisted development.

This systematic process ensures that the MCR is not only statistically based but also aligned with the realities of modern software engineering, where distinguishing between natural and anomalous contributions is critical for accurate performance evaluation.

## 4.2 Mathematical Foundation

This section formalizes the methodology for distinguishing and analyzing contributions in software development using measurable constructs.

The methodology leverages quartile-based filtering and interquartile range (IQR) calculations to minimize the influence of outliers and ensure consistency in contribution analysis.

This mathematical foundation provides structured tools for analyzing contributions, focusing on measurable parameters to classify and rate commits within defined intervals.

### 4.2.1 Commit Time Delta (CTD) and Model CTD Range (MCTDR)

The Commit Time Delta (CTD) represents the time interval between consecutive commits by the same developer. Given a sequence of commits $C_1, C_2, \ldots, C_n$ with timestamps $t_1, t_2, \ldots, t_n$:

$$CTD_i = t_{i+1} - t_i, \quad for\ i = 1, 2, \ldots, n-1$$

The Model CTD Range (MCTDR) is defined as [L,U], where L and U are empirically derived bounds indicating intervals associated with uninterrupted work sessions:

$$L \leq CTD_i \leq U$$

### 4.2.2 Contribution Size and Quartile Segmentation

Each commit $C_i$ has an associated contribution size $S_i$, which quantifies the magnitude of changes (e.g., lines added or modified). The statistical distribution of $S_i$ is segmented into quartiles:

$$IQR = Q_3 - Q_1$$

Where $Q_1$ and $Q_3$ represent the first and third quartiles, respectively. Contributions are identified as outliers and disqualified if:

$$S_1 < Q_1 - 1.5 \times IQR\ or\ S_1 > Q_3 + 1.5 \times IQR$$

### 4.2.3 Mean-High Model Contribution Rate (mhMCR)

The computation of the Mean-High Model Contribution Rate (mhMCR) focuses on contributions classified within the highest quartile of sizes. This process involves the following steps:

1. **Determine Quartiles:** For all contribution rates $C_1, C_2, \ldots, C_n$ within the Model CTD Range (MCTDR), sort the contributions in ascending order and compute the quartiles $Q_1, Q_2, Q_3$ and $Q_4$:

   - $Q_4$: Commits in the top 25% by contribution rate, determined as the values above the third quartile $Q_3$.

2. **Identify Contributions in $Q_4$:** Contributions $C$ classified as $Q_4$ meet the condition:

$$C > Q_3$$

   This subset $C_{Q_4}$ consists of all contributions in the upper quartile of sizes.

3. **Calculate mhMCR:** The mean of the contribution rates in $C_{Q_4}$ is computed as:

$$mhMCR = \frac{Total\ CR\ of\ C_{Q_4}}{Count\ of\ C_{Q_4}}$$

# 5 Experiment Design

To evaluate model contributions within the context of salaried, professional development workflows, three commercial software repositories were selected. These repositories were chosen because they prominently feature patterns aligning with the research methodology, specifically model contributions indicative of continuous, logical development sessions. By using these datasets, the study isolates contributions made under consistent working conditions, minimizing the influence of extraneous variables often found in volunteer or open-source projects.

The empirical analysis compares two methodologies: the Mean Model Contribution Rate (Mean MCR) and the High Mean Model Contribution Rate (High Mean MCR). These methods serve to identify model contributions and derive resulting contribution rates, with the objective of determining the relative accuracy and validity of each approach. The evaluation focuses on the rates of Unlikely Resolved Effort (URE) associated with each method, providing a quantitative measure of their efficacy. Lower URE rates signify a more precise alignment between the contribution rate model and realistic developer effort.

This methodological framework tests the hypothesis that High Mean MCR offers superior performance over Mean MCR.

## 5.1 Dataset Details

### 5.1.1 Commercial Dataset 1

**Commit Time Delta (CTD):** The dataset exhibits a highly variable distribution of CTD values, ranging from a minimum of approximately one second to over two years (19,152.76 hours). The median value of 47.66 hours suggests that typical commits occur within a two-day window. However, the dataset's mean of 477.09 hours and standard deviation of 1,574.99 hours highlight significant outliers or periods of extended inactivity. The interquartile range (IQR) indicates a tighter clustering of CTD values between approximately 3.24 and 395.39 hours, capturing more consistent development rhythms.

**Levenshtein Word Distance:** Contribution sizes span a wide range, from 0 to a maximum of 13,889 words, indicating considerable variation in the scope and scale of code modifications. The median size of 10 units suggests that most contributions are relatively modest, while the mean value of 157.46 units, coupled with a standard deviation of 721.23 units, points to occasional large contributions that skew the average. The IQR (2 to 49 units) reflects a concentration of contributions involving smaller, incremental changes typical of iterative development practices.

**Contribution Rate (WPM):** The contribution rate is characterized by significant variability, ranging from 0 to 1,314.65 units per hour. The median contribution rate of 0.00491 units per hour implies a predominance of low-productivity periods in terms of output per unit time. This is reinforced by the relatively low IQR values (0.000354 to 0.066016 units per hour), suggesting that most contributions are made during periods of limited activity. The high standard deviation of 73.40 units per hour indicates the presence of sporadic high-output events, likely corresponding to bursts of intensive work or automated changes.

*Table 1: Commercial Dataset 1 Statistics*

|  | **CTD Hours** | **Levenshtein Word Distance** | **Contribution Rate (WPM)** |
|---|---|---|---|
| **count** | 557 | 557 | 557 |
| **mean** | 477.088709 | 157.456014 | 8.279693 |
| **min** | 0.000278 | 0 | 0 |
| **25%** | 3.243889 | 2 | 0.000354 |
| **50%** | 47.664444 | 10 | 0.00491 |
| **75%** | 395.388056 | 49 | 0.066016 |
| **max** | 19152.76472 | 13889 | 1314.6536 |

### 5.1.2 Commercial Dataset 2

**Commit Time Delta (CTD):** The dataset reflects a diverse range of CTD values, starting from approximately one second (0.000278 hours) and extending to nearly 1.2 years (10,461.74 hours). A median CTD of 130.52 hours indicates typical commit intervals around five days, while the mean of 416.93 hours and a high standard deviation of 1169.94 hours point to the presence of substantial variability, including periods of sustained activity and inactivity. The interquartile range (21.51 to 401.14 hours) encapsulates a more consistent rhythm of development activities, likely representative of standard workflows.

**Levenshtein Word Distance:** Contributions display significant variability, ranging from no changes (0 units) to an extensive modification of 264,747 units. The median contribution size of 25 units highlights a pattern of smaller, iterative updates as the most frequent behavior. However, the mean value of 5,709.87 units and the standard deviation of 37,155.31 units suggest occasional very large contributions that skew the average, potentially reflecting major refactorings or automated changes. The interquartile range (6 to 102 units) is more indicative of regular, logical contributions, aligning with typical software development efforts.

**Contribution Rate (WPM):** Contribution rates span from 0 to an extraordinary 354,151.34 units per hour, underscoring extreme variability in productivity measures. With a median rate of 0.00503 units per hour, most contributions occur at relatively low productivity levels. The mean contribution rate of 556.04 units per hour, combined with a standard deviation of 12,868.59 units per hour, points to the influence of sporadic high-output contributions. The interquartile range (0.000757 to 0.042216 units per hour) suggests that typical contributions reflect steady but modest activity, aligning with consistent development workflows rather than extreme outliers.

*Table 2: Commercial Dataset 2 Statistics*

|       | CTD Hours   | Levenshtein Word Distance | Contribution Rate (WPM) |
|-------|-------------|---------------------------|-------------------------|
| count | 1301        | 1301                      | 1301                    |
| mean  | 416.929353  | 5709.869331               | 556.037653              |
| min   | 0.000278    | 0                         | 0                       |
| std   | 1169.940952 | 37155.31025               | 12868.58602             |
| 25%   | 21.505556   | 6                         | 0.000757                |
| 50%   | 130.520833  | 25                        | 0.005032                |
| 75%   | 401.141111  | 102                       | 0.042216                |
| max   | 10461.74472 | 264747                    | 354151.34               |

### 5.1.3 Commercial Dataset 3

**Commit Time Delta (CTD):** The dataset shows a wide variability in commit intervals, ranging from approximately 15 seconds (0.004 hours) to nine months (6,480.29 hours). The median CTD of 182.98 hours indicates that typical development sessions involve commits roughly every 7.6 days. The interquartile range (IQR), spanning from 46.20 to 604.30 hours, reflects a broad distribution of more consistent commit patterns. A mean CTD of 555.06 hours, coupled with a standard deviation of 934.58 hours, highlights significant outliers or periods of inactivity that skew the average.

**Levenshtein Word Distance:** Contribution sizes vary from 0 (indicating whitespace only change) to a maximum of 6,376 words, with a median of 38 words, suggesting that most contributions involve modest changes to the codebase. The mean contribution size of 218.01 units and the high standard deviation of 637.61 units suggest the presence of occasional large contributions. The IQR of 6 to 137 units confirms that the majority of contributions are relatively incremental and align with iterative development practices.

**Contribution Rate (WPM):** Contribution rates range from 0 to 2,075.45 units per hour, reflecting significant variation in developer output. The median contribution rate is 0.00334 units per hour, indicating that most contributions occur during periods of low activity. The mean rate of 6.02 units per hour, combined with a high standard deviation of 97.95 units, points to sporadic bursts of high productivity, possibly corresponding to intensive work sessions or automated code generation. The IQR, spanning from 0.000507 to 0.026783 units per hour, suggests that the bulk of contributions are made during periods of moderate activity, typical of steady development workflows.

*Table 3: Commercial Dataset 3 Statistics*

|       | CTD Hours   | Levenshtein Word Distance | Contribution Rate (WPM) |
|-------|-------------|---------------------------|-------------------------|
| count | 901         | 901                       | 901                     |
| mean  | 555.05633   | 218.014428                | 6.023211                |
| min   | 0.004167    | 0                         | 0                       |
| 25%   | 46.201667   | 6                         | 0.000507                |
| 50%   | 182.980833  | 38                        | 0.00334                 |
| 75%   | 604.301111  | 137                       | 0.026783                |
| max   | 6480.291389 | 6376                      | 2075.4546               |

## 5.2 Experiment Steps

This section outlines the experimental methodology used to evaluate contribution rate calculations and imputations in software development data. The process is systematic, leveraging programmatic steps derived from the Python-based data analysis workflows. This sequence ensures a reproducible approach to evaluating contribution rates and their impact on developer productivity metrics.

### 5.2.1 Data Preparation

1. Load datasets of commit metadata, including commit time deltas (CTD) and contribution sizes.
2. Filter out records with missing or invalid data to ensure consistency across the analysis.
3. Convert raw data into a standardized format, ensuring uniformity in time units (e.g., hours) and contribution size metrics.

### 5.2.2 Define Contribution Classes

1. **Model Contributions:** Contributions falling within a predefined CTD range that represents uninterrupted work sessions. These contributions serve as candidates for further analysis.
2. **Anti-Model Contributions:** Contributions outside the CTD range, representing interrupted or irregular work patterns.

### 5.2.3 Contribution Rate Calculation

1. **Compute Raw Contribution Rate:** Divide each contribution's size by its associated CTD. For each commit:

$$Contribution\ Rate = \frac{Contribution\ Size}{CTD}$$

2. **Adjust Contribution Rates:** Remove outliers by applying interquartile range (IQR) filtering to isolate typical contribution rates. Contributions outside $[Q1 - 1.5 \times IQR, Q3 + 1.5 \times IQR]$ are flagged and excluded.

### 5.2.4 Imputation of Contribution Rates

1. Assign imputed contribution rates to Anti-Model Contributions (AMCs) using the median contribution rate of Model Contributions (MCs) within the dataset.
2. Calculate imputed resolved effort hours for each AMC:

$$Resolved\ Effort\ Hours\ (RHE) = Imputed\ Rate\ \times CTD$$

### 5.2.5 Evaluate Unlikely Resolved Effort (URE)

1. Compare each contribution's resolved effort hours (REH) to a practical threshold (e.g., 8 hours/day).
2. Flag contributions exceeding the threshold as exhibiting URE.
3. Record and quantify URE occurrences for each dataset.

### 5.2.6 Method Comparison

1. **Mean Model Contribution Rate (Mean MCR):** Calculate the mean contribution rate across all MCs and use it for imputation.
2. **High Mean Model Contribution Rate (High Mean MCR):** Compute the mean of the top quartile of contribution rates and use this for imputation.
3. Compare the URE rates resulting from each method, noting instances of over- or under-estimated contributions.

### 5.2.7 Result Compilation and Analysis

1. Aggregate results for all datasets, summarizing contribution rate distributions, imputed resolved effort hours, and URE instances.
2. Present findings in tabular and graphical formats to facilitate comparison between Mean MCR and High Mean MCR methods.

## 6 Experiment Results

In this section, we provide a structured analysis of each dataset, focusing on metrics that reveal patterns in contribution behavior. Each dataset is accompanied by the following components:

1. **Chart: "Candidate Model Contributions"**
   This chart visualizes the full distribution of contribution rates, measured in words per minute (WPM), across the temporal scale of commit time deltas (CTD). It displays all contributions within the dataset, illustrating how productivity rates vary with different intervals between commits. This chart serves as the foundation for identifying patterns, anomalies, and trends in developer contributions across the full dataset.

2. **Chart: "IQ4 Model Contributions"**
   This chart is an excerpt of the "CTD Hours by Contribution Rate (WPM)" chart, specifically zooming in on contributions selected as "model contribution rates." These contributions are those classified within the highest quartile (IQR4) of contribution sizes, representing the most productive and consistent work sessions. By isolating this subset, the chart highlights the characteristics of contributions deemed representative of sustained and logical development activity.

3. **Table: "CRIM Metric Details"**
   This table provides a detailed summary of the dataset's key metrics, including:

   - **Mean Contribution Rates (mMCR and mhMCR):** Summarizes both the overall average (mMCR) and high-mean productivity rates (mhMCR), emphasizing differences in modeling approaches.
   - **Commit Counts by Categories:** Breakdowns of commits into categories such as "Quick Remedy Commits," "Model Contribution Commits," and "Disqualified Model Contribution Candidates" (dMCCs).
   - **URE and Imputation Statistics:** Data on "Unlikely Resolved Effort" (URE) commits and imputed contributions, offering insights into the modeling methods' precision.

The **mhMCR over mMCR Improvement Percent** reflects the relative enhancement of the High Mean Model Contribution Rate (mhMCR) method over the Mean Model Contribution Rate (mMCR). This metric is significant in the context of the experiment as it quantifies the extent to which the mhMCR methodology better isolates and emphasizes the most productive contributions, compared to the overall mean approach.

## 6.1 Commercial Dataset 1

*Figure 1: Commercial Dataset 1 Model Contribution Rate Candidates*

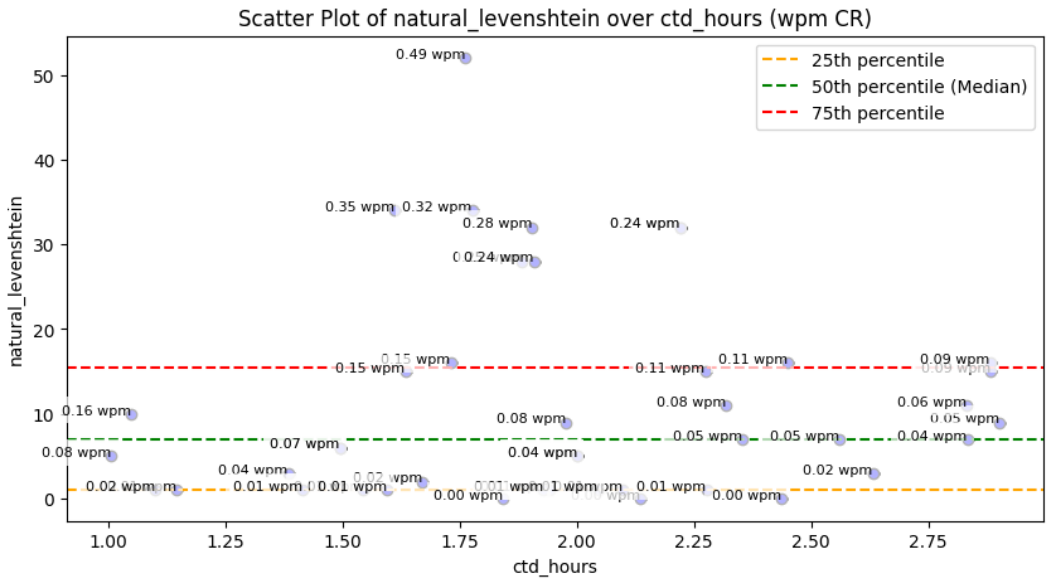

*Figure 2: Commercial Dataset 1 IQ4 Model Contribution Rate Candidates*

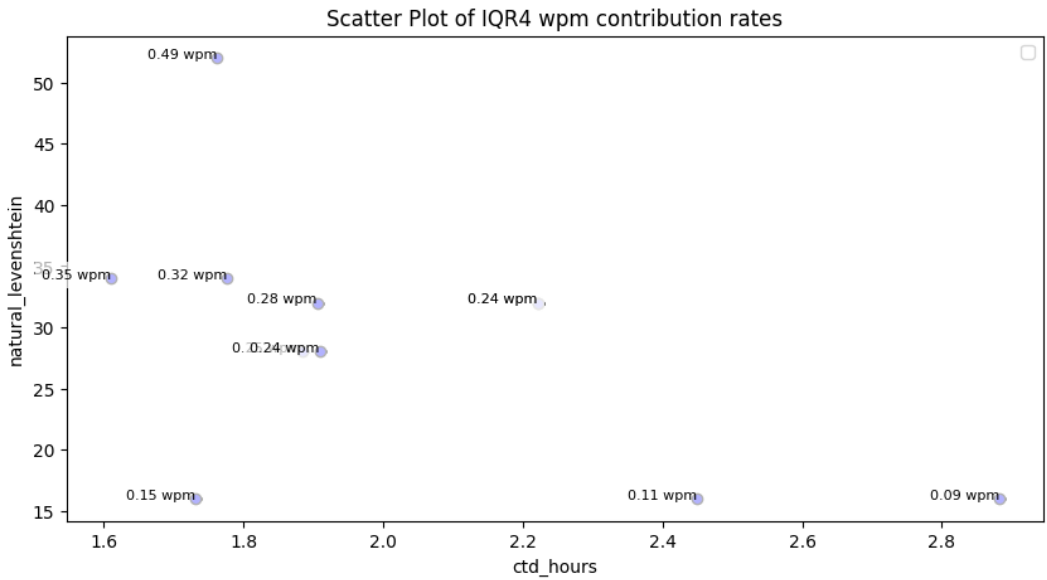

*Table 4: Commercial Dataset 1 CRIM Metrics*

| Metric | Value |
|---|---|
| Mean Model Contribution Rate (wpm) | 0.097 |
| Mean-High Model Contribution Rate (wpm) | 0.252 |
| Count of Commits without a CTD value | 1528 |
| Count of Quick Remedy Commits | 136 |
| Count of Model Contribution Commits | 11 |
| Count of Disqualified Model Contribution Commits | 32 |
| Count of Unbound Commits | 421 |
| Count of Imputed Commits | 1981 |
| Count of Non-Imputed Commits | 147 |
| Count of mhMCR Based URE Commits | 196 |
| Count of mMCR Based URE Commits | 234 |
| **mhMCR over mMCR Improvement Percent** | **16.24** |

## 6.2 Commercial Dataset 2

*Figure 3: Commercial Dataset 2 Model Contribution Rate Candidates*

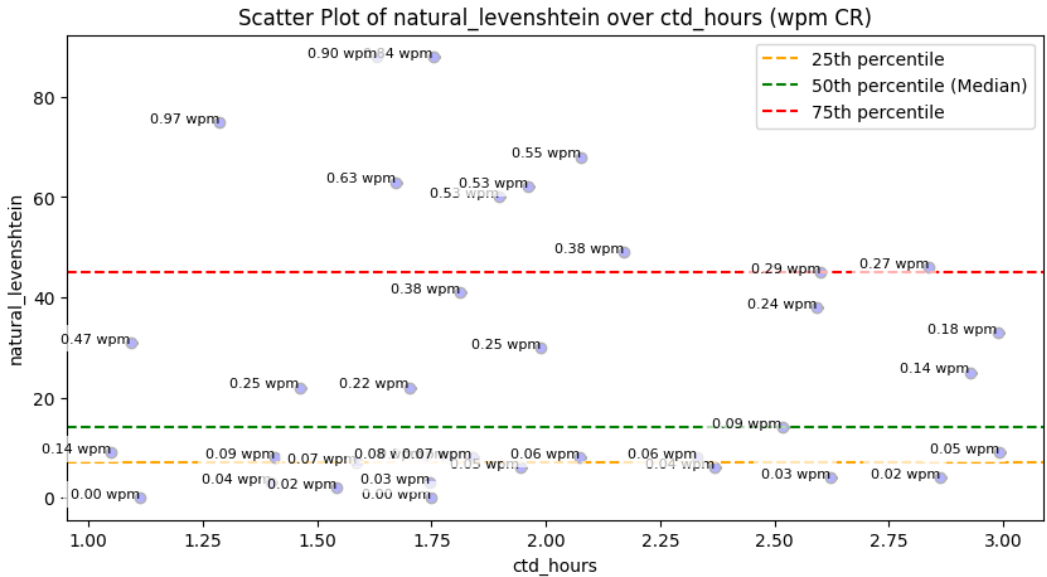

*Figure 4: Commercial Dataset 2 IQ4 Model Contribution Rate Candidates*

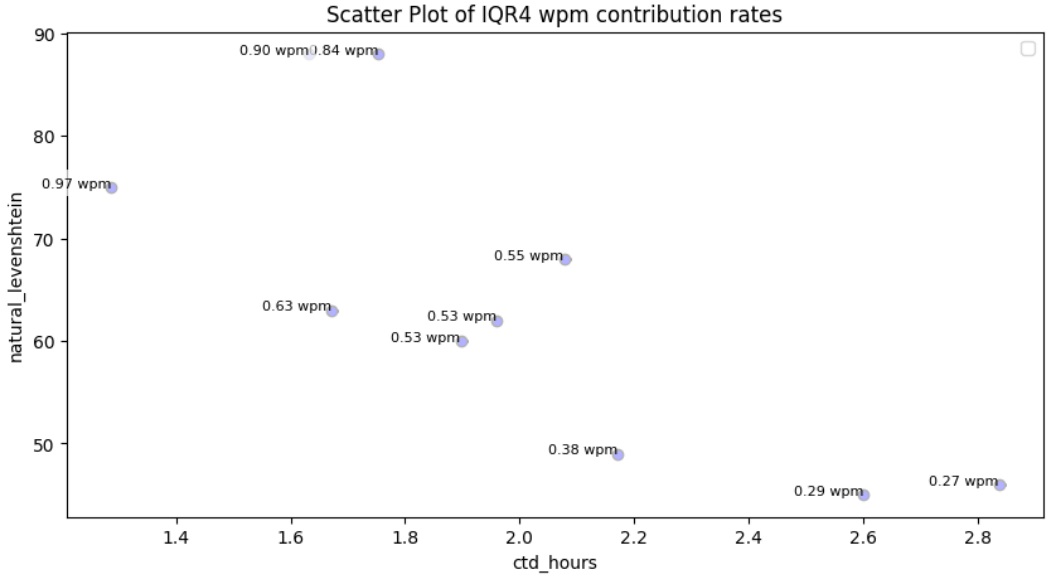

*Table 5: Commercial Dataset 2 CRIM Metrics*

| Metric | Value |
|---|---|
| Mean Model Contribution Rate (wpm) | 0.244 |
| Mean-High Model Contribution Rate (wpm) | 0.587 |
| Count of Commits without a CTD value | 340 |
| Count of Quick Remedy Commits | 182 |
| Count of Model Contribution Commits | 10 |
| Count of Disqualified Model Contribution Commits | 27 |
| Count of Unbound Commits | 1119 |
| Count of Imputed Commits | 1486 |
| Count of Non-Imputed Commits | 192 |
| Count of mhMCR Based URE Commits | 377 |
| Count of mMCR Based URE Commits | 540 |
| **mhMCR over mMCR Improvement Percent** | **30.19** |

## 6.3 Commercial Dataset 3

*Figure 5: Commercial Dataset 3 Model Contribution Rate Candidates*

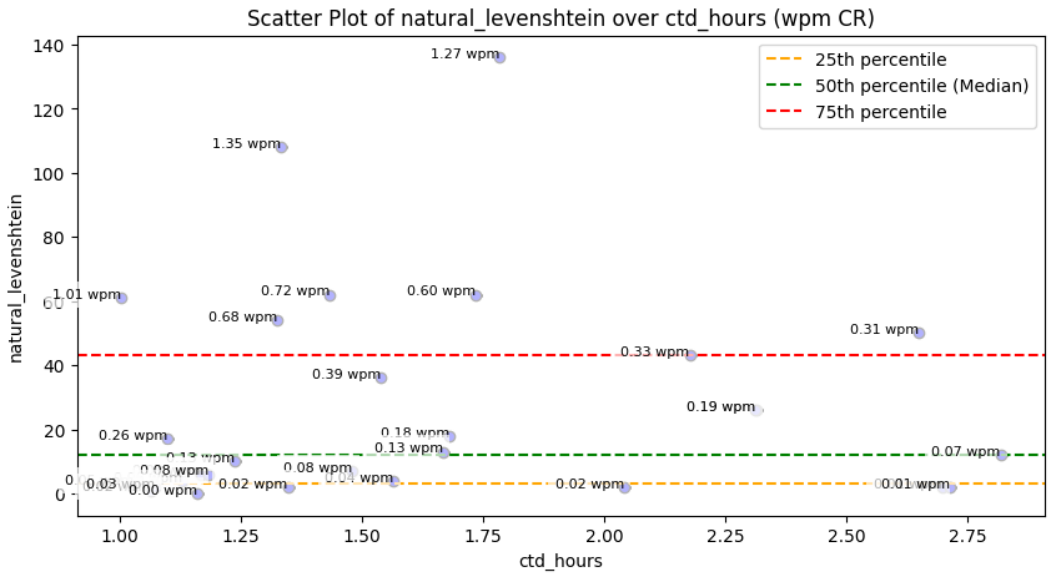

*Figure 6: Commercial Dataset 3 IQ4 Model Contribution Rate Candidates*

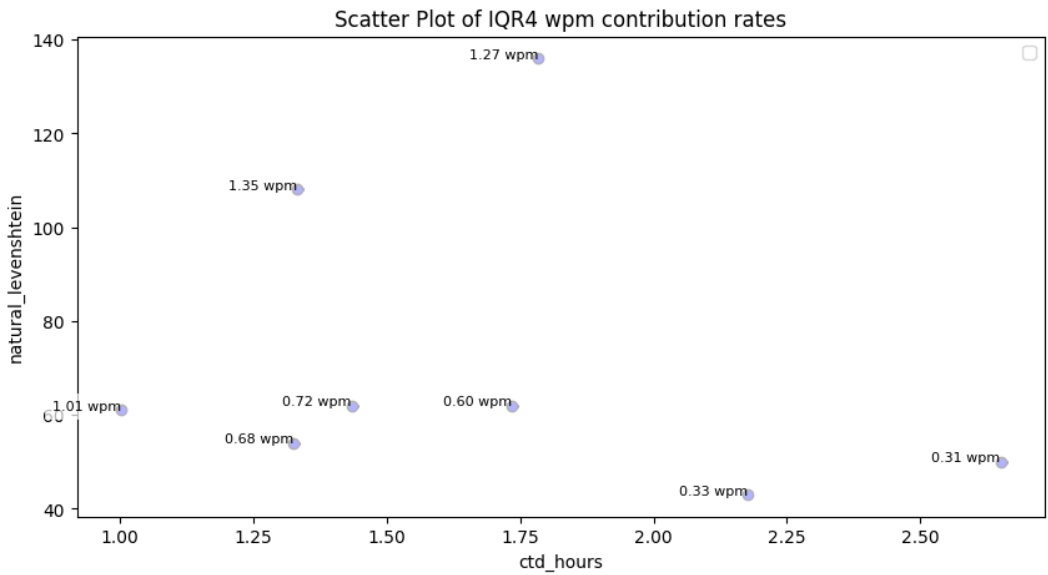

*Table 6: Commercial Dataset 3 CRIM Metrics*

| Metric | Value |
|---|---|
| Mean Model Contribution Rate (wpm) | 0.287 |
| Mean-High Model Contribution Rate (wpm) | 0.784 |
| Count of Commits without a CTD value | 301 |
| Count of Quick Remedy Commits | 93 |
| Count of Model Contribution Commits | 8 |
| Count of Disqualified Model Contribution Commits | 21 |
| Count of Unbound Commits | 808 |
| Count of Imputed Commits | 1130 |
| Count of Non-Imputed Commits | 101 |
| Count of mhMCR Based URE Commits | 227 |
| Count of mMCR Based URE Commits | 333 |
| **mhMCR over mMCR Improvement Percent** | **31.83** |

# 7 Analysis

The **mhMCR over mMCR Improvement Percent** measures the relative enhancement of the High Mean Model Contribution Rate (mhMCR) method compared to the Mean Model Contribution Rate (mMCR) approach. By focusing on the highest quartile of contribution rates, mhMCR emphasizes peak productivity, enabling a more refined analysis of sustained development efforts. This section analyzes the significance of this metric across the datasets and evaluates their contribution patterns.

## 7.1 Dataset 1: Gradual but Distinct Improvements

- **Improvement Percent**: 16.24%
- Dataset 1 contains the widest range of commit time deltas (0.000278 to 19,152.76 hours) and a median CTD of 47.66 hours. The variability indicates a mix of frequent commits and prolonged inactivity, likely reflecting both short-term fixes and long-term project planning phases.
- The **mhMCR value (0.252)**, although modest, demonstrates a significant improvement over **mMCR (0.097)**. This suggests that the highest quartile isolates meaningful contributions better than the overall average in a dataset with high temporal dispersion.
- Observations reveal a concentration of productive commits near the median CTD and contribution size, with low median contribution rates (0.00491 units/hour). This highlights the necessity of excluding lower quartile contributions for precise productivity assessment.

## 7.2 Dataset 2: Largest Improvement and Broadest Range

- **Improvement Percent**: 30.19%
- Dataset 2 exhibits a broader range of contribution sizes (0–264,747 units) and rates (0–354,151.34 units/hour). The median contribution rate (0.00503 units/hour) and CTD (130.52 hours) are higher than Dataset 1, indicating more substantial contributions and longer average intervals between commits.

- The **mhMCR (0.587)** is over double the **mMCR (0.244)**, reflecting the effectiveness of the mhMCR method in a dataset characterized by high variability and significant outliers.
- The dataset's larger improvement percent demonstrates the benefit of emphasizing high-quartile contributions in environments where wide disparities in contribution sizes and rates can obscure overall performance metrics.

### 7.3 Dataset 3: Focused Contributions with Consistent Improvement

- **Improvement Percent**: 31.83%
- Dataset 3 features a narrower range of contribution sizes (0–6,376 units) compared to Dataset 2 but higher median values for both CTD (182.98 hours) and contribution size (38 units). The smaller spread suggests more cohesive work sessions and less variability in commit patterns.
- The **mhMCR (0.784)** outperforms the **mMCR (0.287)** by the largest margin. This improvement highlights the ability of mhMCR to isolate productive contributions, even in datasets where overall contribution sizes are less variable.
- The median contribution rate (0.00334 units/hour) is slightly lower than Dataset 1, but the improvement percent underscores that focusing on high-performing commits consistently yields a more accurate productivity benchmark.

### 7.4 Cross-Dataset Observations

1. **Correlation with Variability**: The datasets with higher variability in CTD and contribution size benefit most from the mhMCR method. Dataset 2's improvement percent is indicative of this trend, as it exhibits the largest disparities.
2. **Productivity Isolation**: Across all datasets, the mhMCR method successfully isolates contributions indicative of sustained and meaningful work. This is evidenced by the significant reduction in Unlikely Resolved Effort (URE) instances when using mhMCR over mMCR.
3. **Consistency of Improvement**: Despite differing dataset characteristics, the improvement percentages remain substantial (16.24%–31.83%), underscoring the robustness of the mhMCR methodology.

### 7.5 Implications of Findings

The **mhMCR over mMCR Improvement Percent** validates the hypothesis that emphasizing peak contributions through the mhMCR method offers a more precise productivity measure. This improvement is particularly evident in datasets with high variability or pronounced outliers. By filtering for the top quartile, the mhMCR method mitigates noise from low-value contributions, aligning productivity metrics with realistic and impactful developer efforts.

This analysis demonstrates the value of mhMCR in software engineering datasets, providing a foundation for further applications, such as team performance assessments and workflow optimization. The significant enhancement across diverse datasets reinforces its utility as a robust metric in empirical software development research.

# 8 Conclusions

## 8.1 Summary of Findings

This study builds upon the "Time Delta Method for Measuring Software Development Contribution Rates" dissertation [2] to expand and formalize the **Model Contribution Rate (MCR) theory**, which isolates natural contributions indicative of sustained, logical developer activity. The research introduces and empirically validates the **Mean-High Model Contribution Rate (mhMCR)** method, a significant refinement of contribution rate imputation that emphasizes peak productivity. Key findings include:

1. **Validation of Model Contribution Rate Theory**:
   - The study formalized the theory by distinguishing model contributions—indicative of continuous and productive work—from anomalies or anti-model contributions caused by automation or delays.
   - Empirical tests demonstrated that focusing on contributions within the Model Commit Time Delta Range (MCTDR) significantly improved the precision of productivity metrics.
2. **Superiority of the mhMCR Method**:
   - The mhMCR method, which calculates the mean of the highest quartile of model contributions, outperformed the simpler mean model contribution rate (mMCR) in all tested scenarios.
   - Empirical results showed that mhMCR reduces **Unlikely Resolved Effort (URE)** occurrences by better filtering irrelevant or inflated contributions. This establishes mhMCR as a candidate approach for productivity analysis in software engineering workflows.

By integrating the mhMCR method into the Contribution Rate Imputation Method (CRIM), this study provides a precise, practical framework for evaluating software development contributions, enabling more accurate measurement and imputation of developer effort.

## 8.2 Opportunities for Future Research

Building on the insights and methodologies presented in this study, several avenues for future exploration arise, directly informed by themes and challenges identified within the research:

1. **Understanding AI-Assisted Contributions**: As discussed, AI-assisted development is increasingly significant, but its impact on productivity remains underexplored. Future research could analyze AI-assisted contributions in depth, assessing how different AI tools and frameworks influence contribution patterns and productivity metrics over time.
2. **Domain-Specific Applications**: While the study has validated mhMCR in commercial software repositories, further validation across domains such as open-source or volunteer-driven projects could explore its generalizability and highlight differences in contribution patterns across contexts.
3. **Longitudinal and Temporal Analysis**: Observing how contribution rates evolve over time—particularly with the adoption of new tools or methodologies—could provide valuable insights into trends in developer productivity and team performance sustainability.
4. **Toolchain Automation and Process Improvements**: Future research could focus on developing practical toolsets that utilize the mhMCR framework to automate aspects of workflow optimization, such as identifying technical debt, streamlining code reviews, or improving resource allocation. As noted in the analysis, integrating these tools into existing pipelines offers significant potential to enhance team productivity.

5. **Refinement of Model Assumptions**: Expanding the theoretical underpinnings of the mhMCR model by exploring alternative clustering techniques or imputation methods could improve the robustness of productivity metrics in environments with high variability or noise.

## 8.3  Final Remarks

This research advances the field of software engineering by presenting a refined, empirically validated framework for contribution rate analysis. The **mhMCR methodology** not only provides a model for measuring developer productivity but also establishes a benchmark for precision in effort imputation. These contributions equip practitioners and researchers with tools to better understand and optimize team performance, bridging the gap between theoretical insights and practical applications in modern development ecosystems.

# Appendix

**Terminology**

| Term | Definition |
|---|---|
| Actual Hours (AH) | Work hours derived directly from CTD as a proxy for time spent on a contribution. |
| Anti-Model Contribution (AMC) | Commits deviating from normal work patterns and require contribution rate imputation to accurately measure temporal effort. |
| Commit Time Delta (CTD) | The time interval between consecutive commits by the same author, used to infer work session patterns. |
| Contribution Rate Imputation Method (CRIM) | A technique to estimate and assign contribution rates for AMC commits by imputing values based MCR. |
| CRIM Estimated Hours (EH) | Work hours inferred using imputed contribution rates for AMC or UbC commits. |
| Disqualified Model Contribution Candidate (dMCC) | Contributions falling within the MCTD but that are disqualified as model contributions (MC) for failing some criteria (e.g. falling outside the IQR4 of contribution rates in the MCTD) |
| Imputed Commit (IC) | A commit where the contribution rate was imputed, and subsequent REH was also calculated. |
| Knowledge Worker Fatigue | A duration where the accuracy or precision of knowledge worker output degrades due to fatigue. |
| Mean Model Contribution Rate (mMCR) | The average contribution rate calculated across all contributions within the Model CTD Range. |
| Mean-High Model Contribution Rate (mhMCR) | The mean of the highest contribution rates within the Model CTD Range, emphasizing peak productivity periods. |
| Model Contribution (MC) | Contributions within the Model CTD Range, representing natural and consistent developer activity. |
| Model Contribution Candidate (MCC) | A potential MC, falling within the MCTDR but requiring further validation against MC. |
| Model Contribution Rate (MCR) | The baseline rate of contributions during consistent and productive work sessions, serving as a benchmark for analysis. This rate can be calculated by different methods (e.g. mMCR or mhMCR) |
| Model CTD Range (MCTDR) | A range of CTD values suggesting continuous and productive work sessions, falling between quick, high-intensity commits and extended ant-model intervals. |
| Natural Contribution (NC) | Code changes representing logical, developer-driven progression of a codebase. |

| Term | Definition |
|---|---|
| Non-Imputed Commit (nIC) | A commit where the contribution rate was not imputed, and the actual CTD value was assigned as REH. |
| Quick-Remedy Commit (QRC) | A commit with very low CTD but comparatively high contribution measure (CR), typically reflecting rapid, isolated fixes rather than continuous development. |
| Resolved Effort Hours (REH) | The computed hours for a contribution, based on contribution rates and corresponding metrics. |
| Unbound Commit (UbC) | A commit with a high CTD and comparatively low contribution measure (CR), indicating a possible interruption or delayed submission of changes. |
| Unlikely Resolved Effort (URE) | REH exceeding plausible limits (e.g., more than 8 hours/day), suggesting an overestimation of contribution effort. |
| Unnatural Contribution (UnC) | Contributions arising from automated, systematic refactoring or copied code that do not reflect logical developer-driven progression of a codebase. |

## Acknowledgements

The deepest gratitude is expressed to the Systems Engineering Department at Colorado State University, Fort Collins, for their invaluable support in making this research possible.

*During the preparation of this work, artificial intelligence was used in the research and writing processes. After using these services, the output was reviewed and edited as needed and full responsibility is taken for its content.*